\documentclass[%
reprint,
superscriptaddress,
showpacs,preprintnumbers,
 amsmath,amssymb,
 aps,
prd,
]{revtex4-1}

\usepackage{graphicx}
\usepackage{dcolumn}
\usepackage{bm}
\usepackage{bbold}
\usepackage{amssymb,amsmath}
\usepackage{hyperref}

\usepackage{color}
\usepackage{amsfonts}
\usepackage{subfigure}
\usepackage{array}

\newcommand{\tr}{\ensuremath{\operatorname{tr}}}

\newcolumntype{L}{>{\centering\arraybackslash}m{3cm}}

\definecolor{bjcol}{rgb}{1,.44,0.13}

\definecolor{blue}{rgb}{0,0,1}

\definecolor{green}{rgb}{0,1,0}

\definecolor{red}{rgb}{1,0,0}

\definecolor{gray}{rgb}{.5,.5,.5}

\definecolor{darkgreen}{rgb}{.0,.5,.0}

\def\Fig#1{Fig.~\ref{#1}} \def\Tab#1{Tab.~\ref{#1}}
 \def\Tab#1{Tab.~\ref{#1}}

\def\Eq#1{Eq.~(\ref{#1})}

\def\eqref#1{(\ref{#1})}

\def\sec#1{Sec.~\ref{#1}}

\def\lA0{{\langle A_0 \rangle}}
\def\bA0{{\bar{A}_0}}

\def\0#1#2{\frac{#1}{#2}}

\graphicspath{{./figures/}{./}}

\begin{document}

\preprint{}

\title{Baryon number probability distribution at finite temperature
}

\author{Ke-xin Sun}
\affiliation{School of Physics, Dalian University of Technology, Dalian, 116024,
  P.R. China}

\author{Rui Wen}
\affiliation{School of Physics, Dalian University of Technology, Dalian, 116024,
  P.R. China}

\author{Wei-jie Fu}
\email{wjfu@dlut.edu.cn}
\affiliation{School of Physics, Dalian University of Technology, Dalian, 116024,
  P.R. China}


\begin{abstract}

The probability distribution of the net baryon number is investigated within the functional renormalization group approach. We find that the Roberge-Weiss periodicity related to the Z(3) symmetry of the gluon fields results directly in that, states of the net baryon number $N_B=N\pm 1/3$ with $N\in\mathbb{Z}$ are prohibited, and only those of $N_B=N$ are possible. By employing the probability distribution of the net baryon number, we also compute the cumulants of the baryon number distribution, which are found to be well consistent with those obtained from the generalized susceptibilities. A question about the relation between the color confinement and the probability distribution of net baryon number is put forward.

\end{abstract}

\pacs{11.30.Rd, 
         11.10.Wx, 
         05.10.Cc, 
         12.38.Mh  
     }                             
\maketitle


\section{Introduction}
\label{sec:intro}

Studies of the QCD phase structure have attracted lots of attentions in recent years. The Beam Energy Scan (BES) program at Relativistic Heavy Ion Collider (RHIC) is aiming at locating the critical end point (CEP) of the QCD \cite{Luo:2017faz}, which separates the first-order phase transition at high densities from the continuous crossover at high temperature in the phase diagram \cite{Stephanov:2007fk}. Interesting and promising results have been arrived at in the experiments, in particular a nonmonotonic behavior of the kurtosis of the net proton number distribution with the variation of the collision energy has been observed \cite{Adamczyk:2013dal,Luo:2015ewa}. 

In accompany with the experimental measurements, theoretical explanations and predictions are indispensable. QCD thermodynamics, such as the equation of state, fluctuations and correlations of conserved charges have been widely studied in lattice simulations, and lots of valuable results have been obtained \cite{Karsch:2001cy,Aoki:2006we,Cheng:2007jq,Borsanyi:2010cj}. Though there is a notorious sign problem at finite densities, recent years have seen significant progresses in the lattice calculations when the chemical potential is not too high \cite{Borsanyi:2014ewa,Ding:2015fca,Bazavov:2017dus,Bazavov:2017tot,Karsch:2017zzw,Borsanyi:2018grb}. In the meantime, nonperturbative continuum functional approaches, such as the Dyson-Schwinger equations \cite{Qin:2010nq,Xin:2014ela,Gao:2016qkh}, the functional renormalization group (FRG) \cite{Braun:2009gm,Herbst:2010rf,Haas:2013qwp,Herbst:2013ail,Herbst:2013ufa,Pawlowski:2014zaa,Skokov:2010wb,Skokov:2010uh,Morita:2013tu,Morita:2014fda,Almasi:2017bhq,Fu:2015naa,Fu:2015amv,Fu:2016tey,Fu:2017vvg,Fu:2018wxq}, etc., are not hampered by this problem, therefore can be employed at large chemical potentials to investigate the QCD phase structure and related phenomenologies.

In this work we will investigate the probability distribution of the net baryon number, which is intimately relevant to the fluctuations of the net proton number, and therefore plays a significant role in the experiments of searching for the CEP of QCD \cite{Luo:2017faz}. We will study the influence of the glue dynamics on the probability distribution, and an interesting interrelation between the Roberge-Weiss periodicity \cite{Roberge:1986mm} due to the Z(3) symmetry of the gluon fields and the resulting probability distribution will be revealed. Furthermore, different from the cumulants of the baryon number distribution, the probability distribution of the baryon number calculated in this work, can be directly used as input for some transport simulations in heavy-ion collisions, see e.g., \cite{Jiang:2015hri,Xu:2016skm}, through which one can assess whether the critical behavior related to the chiral symmetry can survive through the subsequent evolution due to some noncritical effects, for instance, volume corrections, detector acceptance cut, etc. \cite{Braun-Munzinger:2016yjz}.

Calculations in this work are performed within the FRG approach. FRG is a continuum functional field theory, which encodes quantum fluctuations of different wavelengths successively, with the renormalization group (RG) scale running from the ultraviolet (UV) to Infrared (IR) regimes \cite{Wetterich:1992yh}. We refer readers to Refs. \cite{Litim:1998nf, Berges:2000ew,Pawlowski:2005xe,Gies:2006wv,Schaefer:2006sr,Pawlowski:2010ht,Braun:2011pp,vonSmekal:2012vx} for QCD-related reviews and Refs. \cite{Pawlowski:2014aha,Pawlowski:2014zaa,Helmboldt:2014iya,Mitter:2014wpa,Braun:2014ata,Rennecke:2015eba,Fu:2015naa,Fu:2015amv,Wang:2015bky,Cyrol:2016tym,Fu:2016tey,Rennecke:2016tkm,Cyrol:2017ewj,Cyrol:2017qkl,Fu:2017vvg,Cyrol:2018xeq} for recent progresses. 

This paper is organized as follows. In Sec.~\ref{sec:probability} we give a brief introduction about the formalism of the probability distribution. The low energy effective theory and the FRG approach are presented in Sec.~\ref{sec:PQM}. In Sec.~\ref{sec:cumulant} the cumulants of the net baryon number distribution are discussed. We present our numerical results and discussions in Sec.~\ref{sec:num}. Then a summary and conclusion is given in Sec.~\ref{sec:sum}.

\section{The baryon number probability distribution}
\label{sec:probability}

We follow the formalism presented in, e.g., Refs. \cite{Morita:2013tu,Morita:2014fda}, and consider an thermodynamical system with a volume $V$, temperature $T$ and a baryon chemical potential $\mu_B$, which is homogeneous and well equilibrated, the net baryon number $N_B$ probability distribution reads
\begin{align}
  P(N_B;T,V,\mu_B)=&\frac{Z(T,V,N_B)}{\mathcal{Z}(T,V,\mu_B)}\exp\left(\frac{\mu_B N_B}{T}\right)\,,\label{eq:proba}
\end{align}
where $Z(T,V,N_B)$ and $\mathcal{Z}(T,V,\mu_B)$ are the canonical and grand canonical partition functions, respectively. 
Note that, in contradistinction to the ordinary case where net fermion number should be integer, the net baryon number $N_B$ here could be multiples of the baryon number of a quark, i.e., $N_B=N_q/3$ with $N_q\in \mathbb{Z}$ being the net quark number. This is necessary, because if one would like to describe the baryon number probability distribution during the QCD phase transition within one theoretical framework, where the degrees of freedom transform, both quarks and hadrons should be taken into account. Therefore, one arrives at the normalization for the probability distribution as follows
\begin{align}
  &\sum_{N_q=-\infty}^{\infty}P(N_B=N_q/3;T,V,\mu_B)=1\,.\label{eq:norma}
\end{align}
Immediately, upon inserting \Eq{eq:proba},  it follows that 
\begin{align}
  \mathcal{Z}(T,V,\lambda)=& \sum_{N_q=-\infty}^{\infty}Z(T,V,N_B) \lambda^{N_q}\,,\label{}
\end{align}
with the fugacity $\lambda=\exp(\mu_q/T)$ and the quark chemical potential $\mu_q=\mu_B/3$. Obviously, the canonical partition function is just the expansion coefficient of the Laurent series of the grand canonical partition function as a function of fugacity, which leads us to obtain
\begin{align}
  Z(T,V,N_B)=&\frac{1}{2\pi i} \oint_C d \lambda \frac{\mathcal{Z}(T,V,\lambda)}{\lambda^{N_q+1}} \,,\label{}
\end{align}
with $C$ indicating a closed contour with the origin point in it in the complex $\lambda$ plane. Choosing the unit circle for the contour $C$, i.e., $\lambda=e^{i \theta}$, we are led to
\begin{align}
  Z(T,V,N_B=N_q/3)=&\frac{1}{2\pi } \int_{0}^{2\pi } d \theta e^{-i \theta N_q} \mathcal{Z}(T,V,e^{i \theta})\,.\label{eq:Zintegral}
\end{align}
Therefore, the canonical partition function can be deduced from the grand canonical one $\mathcal{Z}(T,V,\mu_q)$ with an imaginary chemical potential, i.e, $\mu_q=i \theta T$. Thus, the task of calculating the net baryon number probability distribution in \Eq{eq:proba} is converted to compute the thermodynamic potential density $\Omega$ with an imaginary chemical potential, which is related to the grand canonical partition function as follows
\begin{align}
  \mathcal{Z}(T,V,\mu_q)=&\exp\left(-\frac{V}{T}\Omega(T,V,\mu_q)\right)\,.\label{eq:Zomega}
\end{align}

\section{The low energy effective theory within the FRG approach}
\label{sec:PQM}

As we have shown in the section above, the thermodynamical potential for a system with an imaginary chemical potential is indispensable to the investigation of the net baryon number probability distribution in \Eq{eq:proba}. In this work, we will focus on two ingredients which affect the baryon number distribution: one is the chiral critical behavior, which is relevant to the chiral phase transition; and the other is the confinement information, that is encoded in the glue dynamics and related to the deconfinement phase transition. It should be emphasized that the glue dynamics plays a significant role in the determination of the probability distribution, as we will show in what follows. This is consistent with the fact that the kurtosis of the baryon number distribution is pronouncedly influenced by the glue dynamics as well, see e.g. \cite{Fu:2015naa,Fu:2018wxq}. In this work we employ the two-flavor Polyakov--quark-meson (PQM) low energy effective model within the FRG approach, for more details see, e.g., \cite{Fu:2015naa,Fu:2015amv,Fu:2016tey}. As a nonperturbative continuum field approach, FRG encodes quantum fluctuations of different wavelengths successively, through the running of RG scale $k$ from the UV to IR regime, and the scale-dependent effective action for the low energy effective model is given by
\begin{align}
  &\Gamma_{k}=\int_{x}\bigg\{Z_{q,k}\bar{q}\big[\gamma_{\mu}\partial_{\mu}-\gamma_{0}(\mu_q+igA_0)\big] q\, +\frac{1}{2}Z_{\phi,k}(\partial_{\mu}\phi)^2\nonumber\\[2ex]
  &+h_{k} \,\bar{q}\left(T^{0}\sigma+i\gamma_{5} \bm{T}\cdot\bm{\pi}\right) q+V_{k}(\rho)-c\sigma+ V_\text{glue}(L,\bar L)\bigg\}\,,\label{eq:action}
\end{align}
with $\int_x=\int_0^{1/T}d x_0 \int d^3 x$. $\phi=(\sigma,\bm{\pi})$ are the mesonic fields, and the effective potential $V_{k}(\rho)$ with $\rho=\phi^2/2$ is $O(4)$ invariant. ($T^{0}$, $\bm{T}$) are the generators in the flavor space with $\mathrm{tr}(T^{i}T^{j})=\frac{1}{2}\delta^{ij}$ and $T^{0}=\frac{1}{\sqrt{2N_{f}}}\mathbb{1}_{N_{f}\times N_{f}}$. $Z_{q,k}$ and $Z_{\phi,k}$ are the anomalous dimensions for the quark and meson, respectively; $h_{k}$ is the Yukawa coupling; the term $-c\sigma$ in \Eq{eq:action} breaks the chiral symmetry explicitly. Moreover, in order to take into account the deconfinement information, we include the temporal component of the gluon background field $A_0$  in the effective action. It can also be transformed into the formalism of the traced Polyakov loop $L$ and its conjugate $\bar L$, i.e.,
\begin{align}
  L(\bm{x})=\frac{1}{N_c} \langle \tr\, \mathcal{P}(\bm x)\rangle\,,\quad  \bar L (\bm x)=\frac {1}{N_c} \langle \tr\,\mathcal{P}^{\dagger}(\bm x)\rangle \,,\label{eq:Lloop}
\end{align}
with
\begin{align}
  \mathcal{P}(\bm x)= \mathcal{P}\exp\bigg(ig\int_0^{\beta}d\tau A_0(\bm{x},\tau)\bigg)\,,\label{eq:Ploop}
\end{align}
where $\mathcal{P}$ on the r.h.s. is the path-ordering. $V_\text{glue}$ in the effective action in \Eq{eq:action} is the glue potential which will be specified in the following.

The evolution equation for the scale-dependent effective action in \Eq{eq:action}, i.e., the Wetterich equation \cite{Wetterich:1992yh}, reads
\begin{align}
  \partial_{t}\Gamma_{k}&=-\mathrm{Tr}\big(G^{q\bar q}_{k}\partial_{t} R^{q}_{k}\big)+\frac{1}{2}\mathrm{Tr}\big(G^{\phi\phi}_{k}\partial_{t} R^{\phi}_{k}\big)\,,
 \label{eq:WetterichEqPQM}
\end{align}
with $t=\ln (k/\Lambda)$ and $\Lambda$ is the initial evolution scale, i.e., the UV cutoff scale. $G_{k}$'s are the propagators for the quark and meson fields; $R_k$'s are the IR regulators that suppress specific quantum fluctuations whose wavelengths are larger than $1/k$. In this work, the local potential approximation (LPA), i.e., $Z_{q,k}=Z_{\phi,k}=1$ and $\partial_t h_{k}=0$, is adopted. Then following from the Wetterich equation,  one immediately obtains the flow equation for the effective potential in \Eq{eq:action}, i.e.,
\begin{align} 
  \partial_{t}V_{k}(\rho)&=\frac{k^{4}}{12\pi^{2}}\bigg[\frac{N_{f}^{2}-1}{\sqrt{1+\bar{m}_{\pi,k}^{2}}}\Big(1+2\,n_B(\bar{m}_{\pi,k}^{2};T)\Big)\nonumber\\[2ex]
  &\quad+\frac{1}{\sqrt{1+\bar{m}_{\sigma,k}^{2}}}\Big(1+2\,n_B(\bar{m}_{\sigma,k}^{2};T)\Big)\nonumber\\[2ex]
  &\quad-\frac{4N_{c}N_{f}}{\sqrt{1+\bar{m}_{q,k}^{2}}}\Big(1-n_f(\bar{m}_{q,k}^{2};T,\mu_q,L,\bar L)\nonumber\\[2ex]
  &\quad-\bar n_f(\bar{m}_{q,k}^{2};T,\mu_q,L,\bar L)\Big)\bigg]\,,\label{eq:Vflow}
\end{align}
where the dimensionless masses for the meson and quark read
\begin{align} 
  \bar{m}_{\pi,k}^{2}&=\frac{V_{k}^{\prime}(\rho)}{k^{2}}\,,\nonumber\\[2ex] 
  \bar{m}_{\sigma,k}^{2}&=\frac{V_{k}^{\prime}(\rho)+2\rho V_{k}^{\prime\prime}(\rho)}{k^{2}}\,,\nonumber\\[2ex]
  \bar{m}_{q,k}^{2}&=\frac{h_{k}^{2}\rho}{2k^{2}}\,. \label{eq:mass} 
\end{align}
The bosonic distribution function is given by
\begin{align}
  n_B(\bar{m}_{\phi,k}^{2};T)=\frac{1}{\exp\Big(k\sqrt{1+\bar{m}_{\phi,k}^{2}}\Big/T\Big)-1}\,.\label{}
\end{align}
When the chemical potential is imaginary, such as $\mu_q=i \theta T$ as discussed in \sec{sec:probability}, the Polyakov loops $L$ and $\bar L$ in \Eq{eq:Lloop} are complex conjugate to each other exactly. Note, however, that a difficulty will arise when the chemical potential is real valued, for example, when we calculate the grand canonical partition function in the denominator in \Eq{eq:proba} with a real-valued $\mu_B$. 

Inserting $\mu_q=i \theta T$, $L=|L| e^{i\phi}$, $\bar L=|L| e^{-i\phi}$, one obtains the fermionic distribution functions as follow
\begin{align} 
 &n_f(\bar{m}_{q,k}^{2};T,\mu_q,L,\bar L)\nonumber\\[2ex]
 =&\Big[1+2|L|e^{E_{q,k}/T}e^{-i(\theta+\phi)}+|L|e^{2E_{q,k}/T}e^{i(\phi-2\theta)}\Big]\nonumber\\[2ex]
   &\bigg/\Big[1+3|L|e^{E_{q,k}/T}e^{-i(\theta+\phi)}+3|L|e^{2E_{q,k}/T}e^{i(\phi-2\theta)}\nonumber\\[2ex]
   &+e^{3E_{q,k}/T}e^{-3i\theta}\Big]\,, \label{eq:mass} 
\end{align}
and 
\begin{align}
  \bar n_f(\bar{m}_{q,k}^{2};T,\mu_q,L,\bar L)=n_f^*(\bar{m}_{q,k}^{2};T,\mu_q,L,\bar L)\,,\label{}
\end{align}
with $E_{q,k}=k\sqrt{1+\bar{m}_{q,k}^{2}}$.

The thermodynamic potential density is given by 
\begin{align}
  \Omega=V_{k=0}(\rho)-c\sigma+V_\text{glue}(L,\bar L)\,,\label{eq:thermopoten}
\end{align}
where the expectation value of the meson field is determined through its equation of motion (EoM), and the normalization $\Omega=0$ at vacuum is assumed. Since the glue potential $V_\text{glue}$ has Z(3) symmetry, the symmetry of the thermodynamic potential in \Eq{eq:thermopoten} with an imaginary chemical potential has been well known, see e.g., \cite{Roberge:1986mm,Sakai:2008py}. We introduce modified Polyakov loops as follow
\begin{align}
  L^{\prime}&=|L| e^{i\phi^{\prime}},\quad \text{and}\quad \bar L^{\prime}=|L| e^{-i\phi^{\prime}}\,,\label{eq:Ploopmodi}
\end{align}
with $\phi^{\prime}=\phi+\theta$. Substituting the modified Polyakov loops above into \Eq{eq:thermopoten}, one immediately recognizes that $ \Omega$ is invariant with the replacement $\theta \rightarrow \theta + \frac{2\pi}{3}$, which is also known as the Roberge-Weiss periodicity. The periodicity of the thermodynamic potential leads to
\begin{align}
  L^{\prime}(\theta)&=L^{\prime}(\theta+\frac{2\pi}{3}),\quad \text{and}\quad \bar L^{\prime}(\theta)&=\bar L^{\prime}(\theta+\frac{2\pi}{3})\,\label{}
\end{align}
as well. Furthermore, it has been found that $L^{\prime}(\theta)$ is not continuous at $\theta=(2n+1)\pi/3$ with $n\in\mathbb{Z}$, when the temperature is above some critical value, which is also called as the Roberge-Weiss phase transition \cite{Roberge:1986mm}.

the Roberge-Weiss periodicity of the thermodynamic potential, leads directly to the same periodicity of the grand canonical partition function with an imaginary chemical potential $\mu_q=i \theta T$ in \Eq{eq:Zomega}. Then we can rewrite \Eq{eq:Zintegral} as follows
\begin{align}
  Z(T,V,N_q)=&\frac{1}{2\pi } \int_{0}^{\frac{2\pi}{3}} d \theta e^{-i \theta N_q}\mathcal{Z}(T,V,\theta) \Big(1+e^{-i \frac{2\pi}{3} N_q}\nonumber\\[2ex]
   &+e^{-i \frac{4\pi}{3} N_q}\Big)\,,\label{}
\end{align}
where the expression in the parentheses in the integrand has an interesting property, i.e., it is nonvanishing, only when the remainder of $N_q$ over 3 is zero. Therefore, the probability to find a system in a state with non-integer baryon numbers is zero, which is apparently a  manifestation of the color confinement.

\section{Cumulants of the baryon number distribution}
\label{sec:cumulant}

With the probability distribution in \Eq{eq:proba} in hand, one can immediately obtain the statistical average for a quantity, denoted here symbolically as $\mathcal{O}(N_B)$, through the equation as follows
\begin{align}
  \langle\mathcal{O}\rangle&=\sum_{N_B=-\infty}^{\infty}\mathcal{O}(N_B)P(N_B)\,,\label{eq:Oaver}
\end{align}
and the $n$-th order cumulant of the baryon number distribution corresponds to $\mathcal{O}=(\delta N_B)^n$ with $\delta N_B=N_B-\langle N_B\rangle$. 

In fact, the cumulants can also be calculated in another commonly employed method, which resorts to derivatives of the thermodynamic potential with respect to the real-valued baryon chemical potential, i.e., the generalized susceptibilities, see e.g., \cite{Fu:2015naa}. The relevant definition reads
\begin{align}
   \chi_n^{\mathrm{B}}&=\frac{\partial^n}{\partial (\mu_{\mathrm{B}}/T)^n}\bigg(-\frac{\Omega}{T^4}\bigg)\,.\label{eq:suscep}
\end{align}
The relations between $\chi_n$ and the cumulants are given by
\begin{align}
  \chi_2^{\mathrm{B}}&=\frac{1}{VT^3}\langle(\delta N_{\mathrm{B}})^2\rangle\,,\nonumber\\[2ex]
  \chi_4^{\mathrm{B}}&=\frac{1}{VT^3}\Big(\langle(\delta N_{\mathrm{B}})^4\rangle-3\langle(\delta N_{\mathrm{B}})^2\rangle^2\Big)\,,
\end{align}
taking the quadratic and quartic orders for instance.

In the following, we will compare results of the cumulants calculated in these two different approaches. The comparison is nontrivial, especially when the glue dynamics and, thus the information of confinement, is encoded in the calculation. This is because when the baryon chemical potential is real valued, the Polyakov loop $L$ and its conjugate $\bar L$ are ill-defined, since they are not complex conjugate to each other. One method commonly employed to bypass this difficulty is to ignore the phase of  $L$ and $\bar L$, and treat them as two different real quantities, which are determined through their respective equations of motion. 

\section{Numerical results}
\label{sec:num}

%
\begin{figure}[t]
\includegraphics[width=0.45\textwidth]{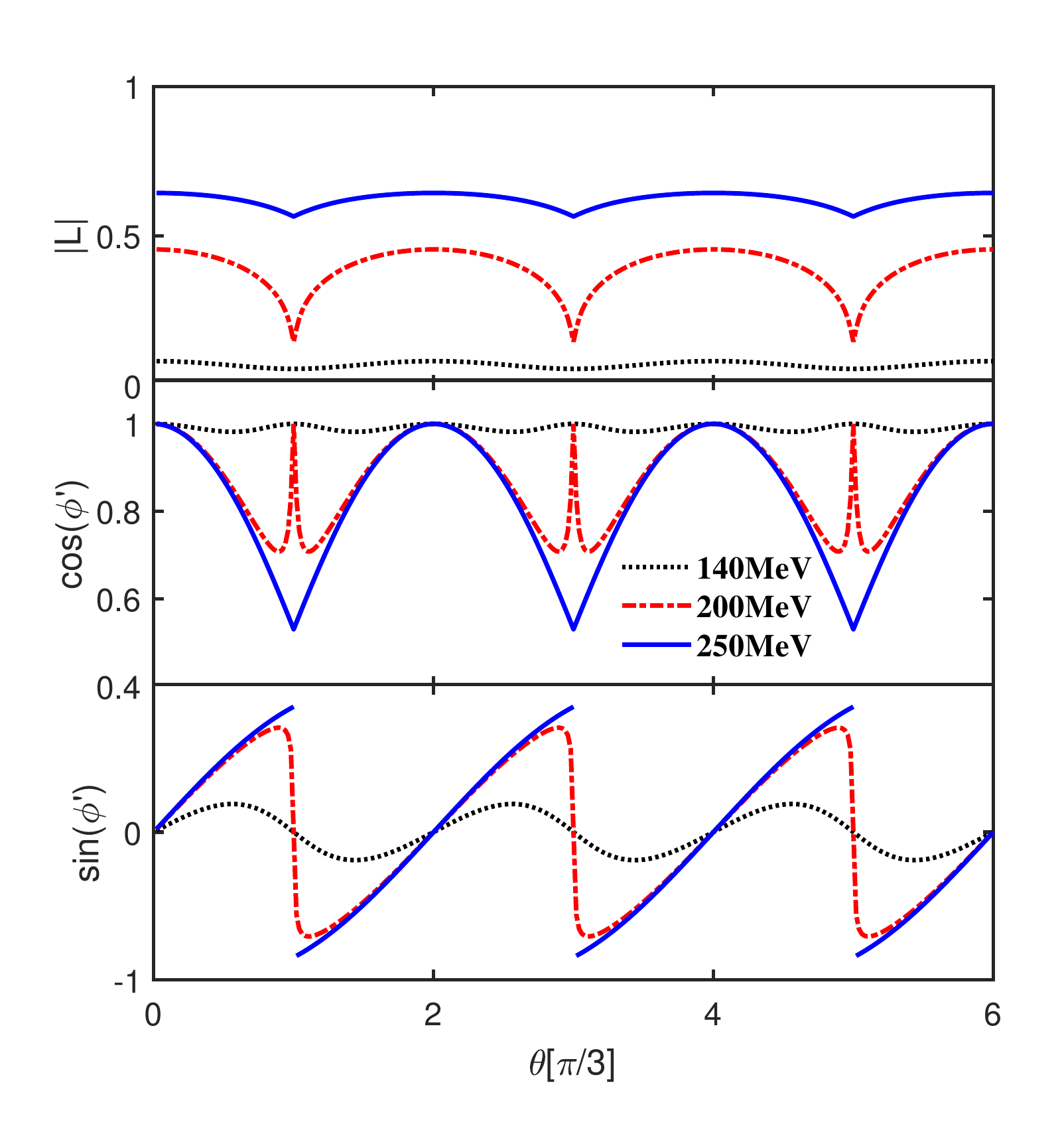}
\caption{Magnitude (top), cosine (middle), and sine (bottom) of the phase of the modified Polyakov loop $L^{\prime}$ in \Eq{eq:Ploopmodi} as functions of the imaginary chemical potential related $\theta$. The dotted, dashed, and solid lines correspond to temperature $T=140$, 200, 250 MeV, respectively.}\label{fig:polyakov}
\end{figure}
%

%
\begin{figure*}[t]
\includegraphics[width=1.0\textwidth]{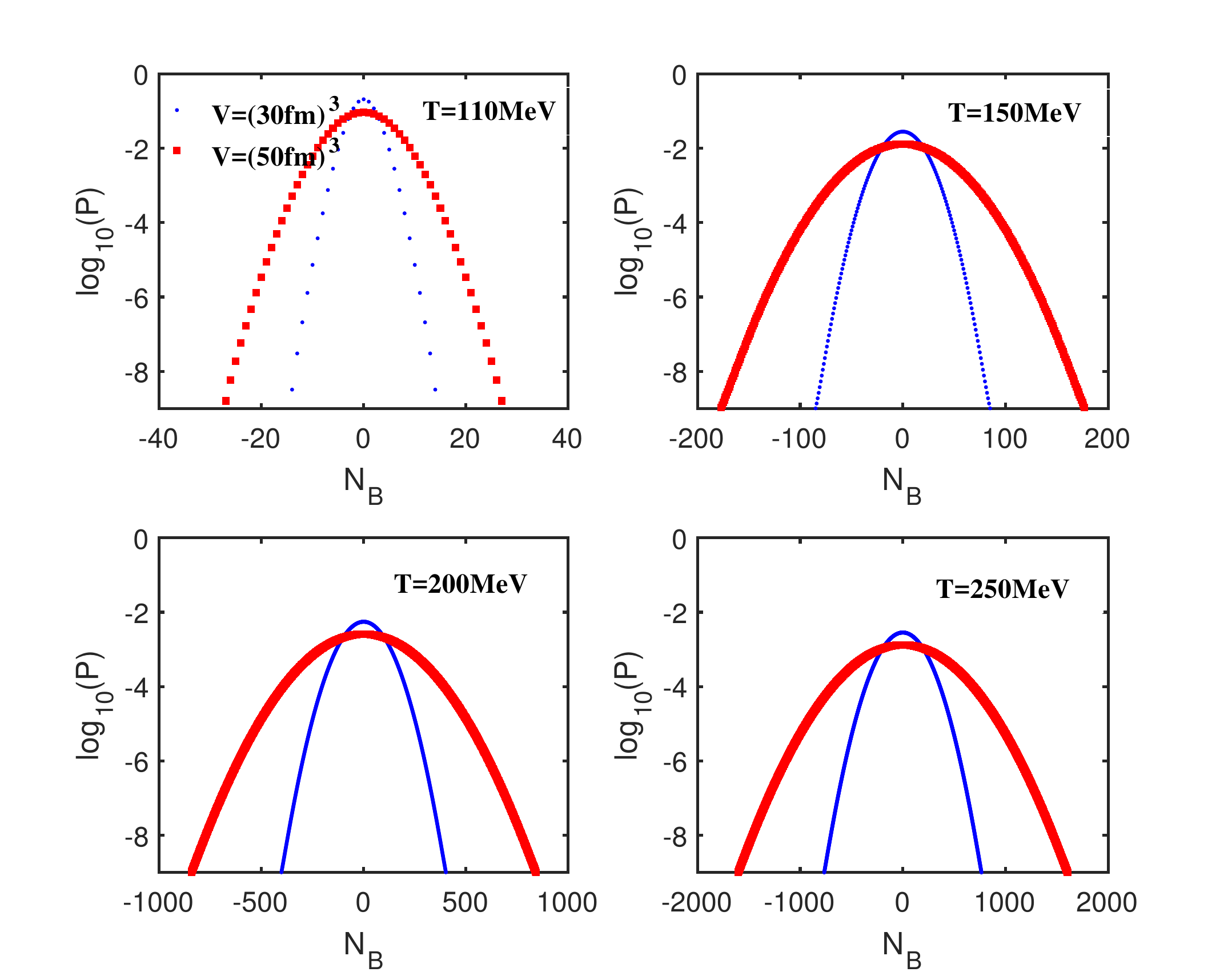}
\caption{Probability distributions of the net baryon number with vanishing real-valued baryon chemical potential, i.e., $\mu_B=0$. Different panels show results corresponding to different values of the temperature. Here we choose two representative volume $V=(30\, \text{fm})^3$ and $(50\, \text{fm})^3$.}\label{fig:PN}
\end{figure*}
%

We employ the Taylor expansion around the physical point to solve the flow equation for the effective potential in \Eq{eq:Vflow}, which reads
\begin{align}
  V_{k}(\rho)&=\sum_{n=0}^{N}\frac{\lambda_{n,k}}{n!}(\rho-\kappa_k)^n\,. \label{eq:VbarTaylor}
\end{align}
Here $\kappa_k$ is the solution of its EoM for every RG scale $k$, which fulfills
\begin{align}
  \frac{\partial}{\partial \rho}\Big(V_{k}(\rho)-c \sigma \Big)\Bigg \vert_{\rho=\kappa_k}&=0\,. \label{eq:Vstat}
\end{align}
Inserting \Eq{eq:VbarTaylor} into \Eq{eq:Vflow} and constraining the expansion point with \Eq{eq:Vstat}, one obtains the flow equation for $\kappa_k$, i.e.,
\begin{align}
  \partial_t \kappa_k=&-\frac{c^2}{\lambda_{1,k}^3+c^2 \lambda_{2,k}}\bigg[\partial_{\rho}\big(\partial_{t}\big|_{\rho} V_{k}(\rho)\big)\Big |_{\rho=\kappa_k}\bigg]\,,\label{}
\end{align}
and those for the expansion coefficients $\lambda_{n,k}$'s, which reads
\begin{align}
  \partial_t \lambda_{0,k}=&\big(\partial_{t}\big|_{\rho}V_{k}(\rho)\big)\Big |_{\rho=\kappa_k}+\lambda_{1,k}\partial_t \kappa_k\,,\\[2ex]
  \partial_t \lambda_{1,k}=&\frac{\lambda_{1,k}^3}{\lambda_{1,k}^3+ c_k^2  \lambda_{2,k}}\bigg[\partial_{\rho}\big(\partial_{t}\big|_{\rho} V_{k}( \rho)\big)\Big |_{\rho= \kappa_k}\bigg]\,,\label{}
\end{align}
and
\begin{align}
  \partial_t \lambda_{n,k}=&\partial_{\rho}^n\big(\partial_{t}\big|_{\rho} V_{k}( \rho)\big)\Big |_{\rho= \kappa_k}+\lambda_{n+1,k}\partial_t  \kappa_k\; (n\ge 2)\,, \label{}
\end{align}
where $\partial_{\rho}^n$ denotes the $n$-th order partial derivative with respect to $\rho$, and $\big|_{\rho}$ the differentiation with $\rho$ fixed. In our calculations, the maximal order of the Taylor expansion $N$ in \Eq{eq:VbarTaylor} is chosen to be $N=5$, which has guaranteed the required convergence very well, for more discussions about the convergence, see e.g. \cite{Pawlowski:2014zaa} .

We evolve these flow equations from an initial UV scale, which is chosen to be $\Lambda=700\,\mathrm{MeV}$ in this work. The effective potential at the initial scale is chiral symmetric, except the explicit breaking induced by the $-c\sigma$ term. Hence, the effective potential at $k=\Lambda$ can be approximated as
\begin{align}
  V_{\Lambda}(\rho)=&\frac{\lambda_{\Lambda}}{2}\rho^2+\nu_{\Lambda}\rho\,.\label{eq:VLambda}
\end{align}
Parameters in the effective potential above, to wit, $\lambda_{\Lambda}$ and $\nu_{\Lambda}$, as well as the Yukawa coupling $h$ and the explicit chiral symmetry breaking related coefficient $c$, are fixed by fitting hadronic observables: the $\pi$- and $\sigma$- mesons masses $m_{\pi}=135.9\,\mathrm{MeV}$ and $m_{\sigma}=502.2\,\mathrm{MeV}$, the $\pi$-meson decay constant $f_\pi =92.0\,\mathrm{MeV}$, and the constituent quark mass $m_{q}=299.1\,\mathrm{MeV}$. Their values are given by $\lambda_{\Lambda}=5.7$, $\nu_{\Lambda}=0.23\,\mathrm{GeV}^2$, $h=6.5$, $c=1.7\times 10^{-3}\mathrm{GeV}^3$, respectively.

%
\begin{figure}[t]
\includegraphics[width=0.5\textwidth]{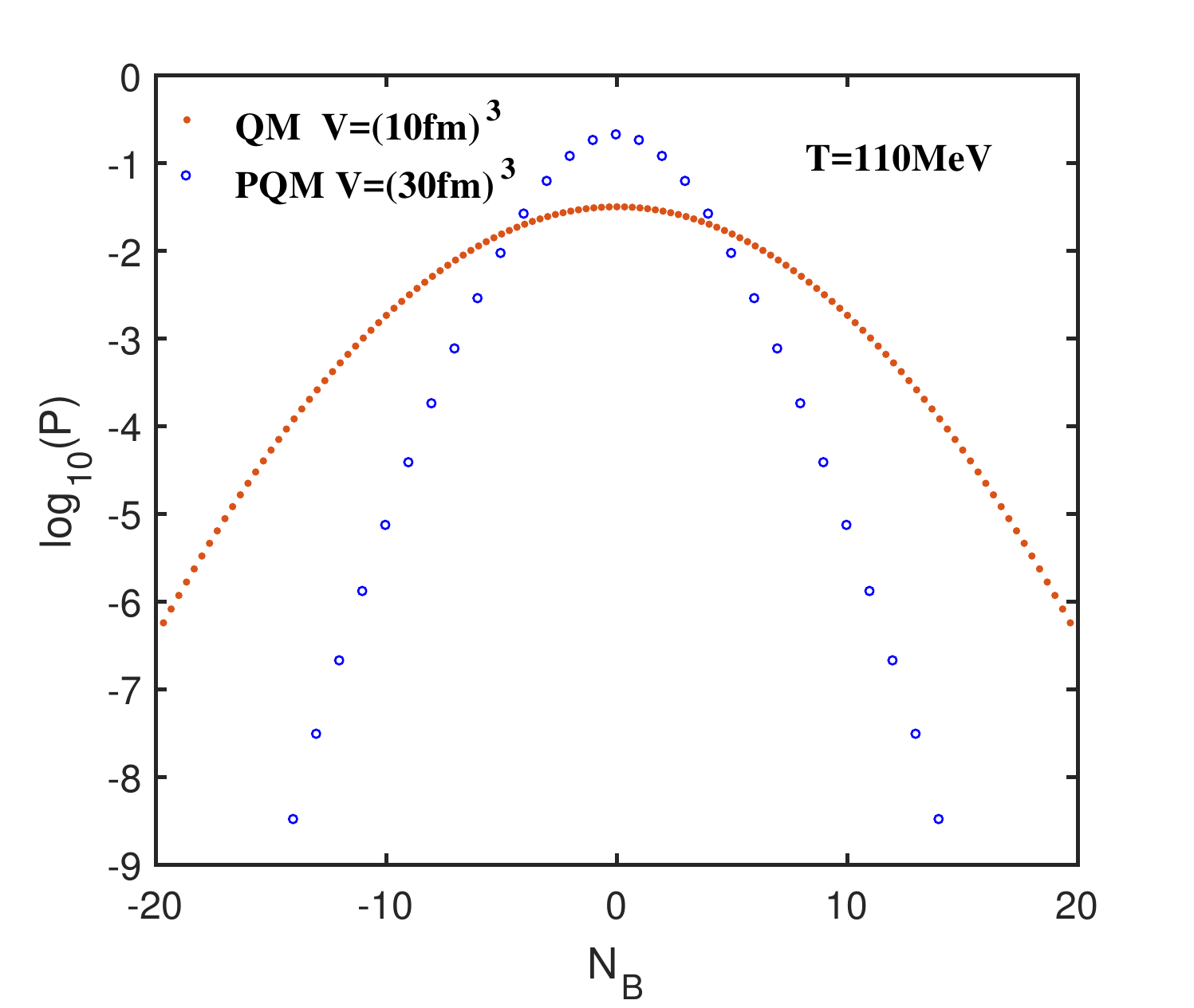}
\caption{Comparison between the probability distribution calculated in the PQM and that in the QM.}\label{fig:compaQM}
\end{figure}
%

It is left to specify the glue potential in Eqs.~(\ref{eq:action}) and ~(\ref{eq:thermopoten}). In this work, we adopt the glue potential parameterized firstly by Lo {\it et al.} in Ref. \cite{Lo:2013hla}, which has also been used in Ref.~\cite{Fu:2016tey}. This potential is characteristic of its ability to describe the quadratic fluctuations of the Polyakov loop, which reads
\begin{align}
  V_\text{glue}(L,\bar{L}) &= -\frac{a(T)}{2} \bar L L + b(T)\ln M_H(L,\bar{L})\nonumber \\[2ex]
  &\quad + \frac{c(T)}{2} (L^3+\bar L^3) + d(T) (\bar{L} L)^2\,,\label{eq:polpot}
\end{align}
with
\begin{align}
M_H (L, \bar{L})&= 1 -6 \bar{L}L + 4 (L^3+\bar{L}^3) - 3  (\bar{L}L)^2\,,
\end{align}
where $M_H$ is the Haar measure of the $SU(3)$ group in color space. The parameterization of the coefficient $a(T) $ is given by
\begin{align}
  a(T) &= \frac{a_1 + a_2/t_{\text{\tiny{YM}}} + a_3/t_{\text{\tiny{YM}}}^2}{1 + a_4/t_{\text{\tiny{YM}}} + a_5/t_{\text{\tiny{YM}}}^2}\,,\label{eq:aT}
\end{align}
and similar parameterizations also apply to $c(T)$ and $d(T)$, but for $b(T)$ it reads
\begin{align}
  b(T ) &= b_1 t_{\text{\tiny{YM}}}^{-b_4} (1 -e^{b_2/t_{\text{\tiny{YM}}}^{b_3}} )\,.\label{eq:bT}
\end{align}
All the constants required for this parameterized glue potential can be found in \cite{Lo:2013hla,Fu:2016tey}. We collect them once more in \Tab{tab:coeffs} for the convenience and completeness. $t_{\text{\tiny{YM}}}$ in Eqs. (\ref{eq:aT}) and (\ref{eq:bT}) are the reduced temperature for the Yang-Mills gauge theory at finite temperature. It has been found that the QCD glue potential, which encodes the backreaction of the matter on the glue sector, can be remarkably well parameterized through the Yang-Mills one \cite{Pawlowski:2010ht,Haas:2013qwp,Herbst:2013ufa}, just by making the following replacement:
\begin{align}
  t_{\text{\tiny{YM}}}&\rightarrow \alpha\,t_{\text{\tiny{glue}}}\quad \text{with}\quad t_{\text{\tiny{glue}}}=(T-T_c^\text{\tiny{glue}})/T_c^\text{\tiny{glue}},\label{}
\end{align}
where the scaling constant $\alpha=0.57$ is found for the two-flavor QCD, and we adopt the critical temperature $T_c^\text{\tiny{glue}}=250\,\mathrm{MeV}$ for the glue potential.
 
%
\begin{table}[tb!]
  \centering
  \begin{tabular}{c||c|c|c|c|c}
     & 1 & 2 & 3 & 4 & 5 \rule{0pt}{2.6ex}\rule[-1.2ex]{0pt}{0pt}\\ \hline\hline
    $a_i$ &-44.14& 151.4 & -90.0677 &2.77173 &3.56403 \\\hline
    $b_i$ &-0.32665 &-82.9823 &3.0 &5.85559  &\\\hline
    $c_i$ &-50.7961 &114.038 &-89.4596 &3.08718 &6.72812\\\hline
    $d_i$ & 27.0885 &-56.0859 &71.2225 &2.9715 &6.61433\\
  \end{tabular}
  \caption{Constants for the parameterization of the glue potential.} 
  \label{tab:coeffs}
\end{table}
%

%
\begin{figure*}[t]
\includegraphics[width=1.\textwidth]{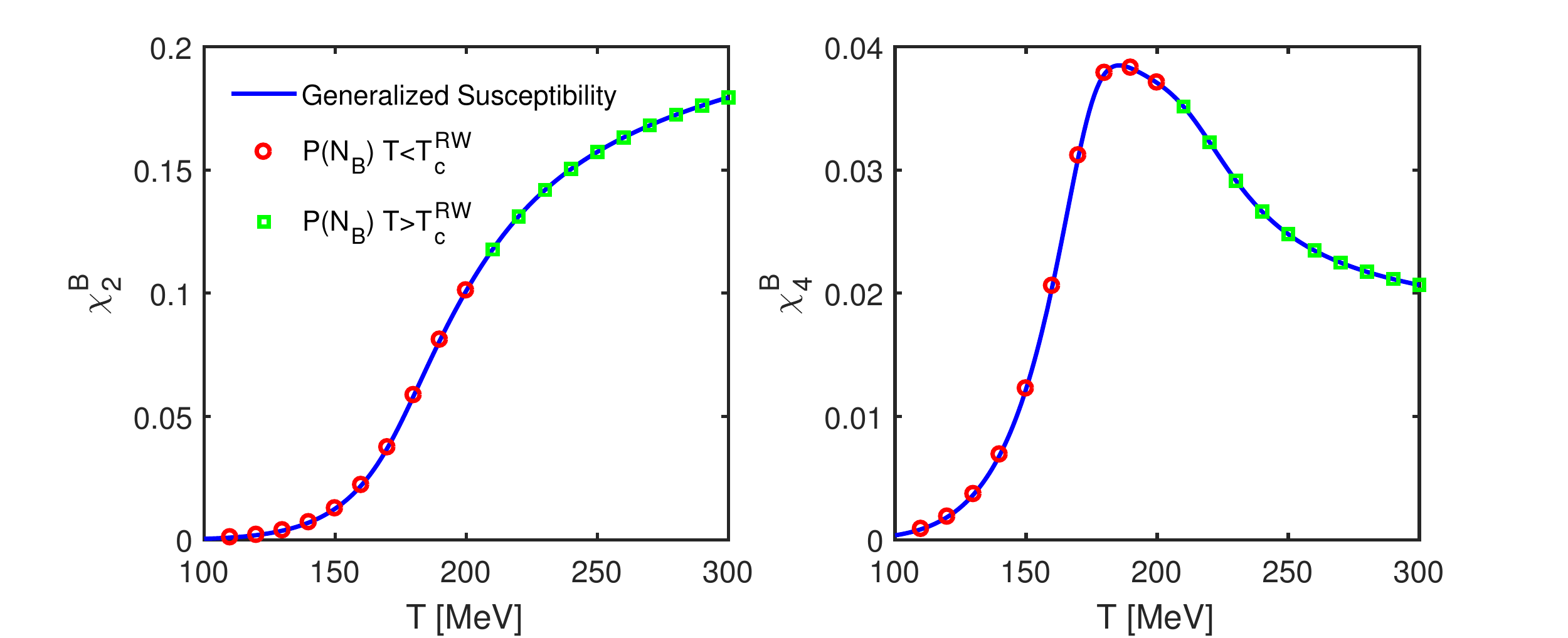}
\caption{Quadratic (left panel) and quartic (right panel) cumulants of the baryon number as functions of the temperature with $\mu_B=0$. Two approaches employed in the calculations are compared. The solid lines correspond to the conventional method, i.e., the generalized susceptibilities in \Eq{eq:suscep},  and the scattering symbols are related to results obtained with the probability distributions of the net baryon number through the equation (\ref{eq:Oaver}). The red open circles correspond to results of temperature below $T_c^{RW}=208\,\text{MeV}$ where there is no discontinuity in the Polyakov loop, while the green open squares to those of temperature above $T_c^{RW}$ with discontinuity.}\label{fig:chi}
\end{figure*}
%

%
\begin{figure}[t]
\includegraphics[width=0.45\textwidth]{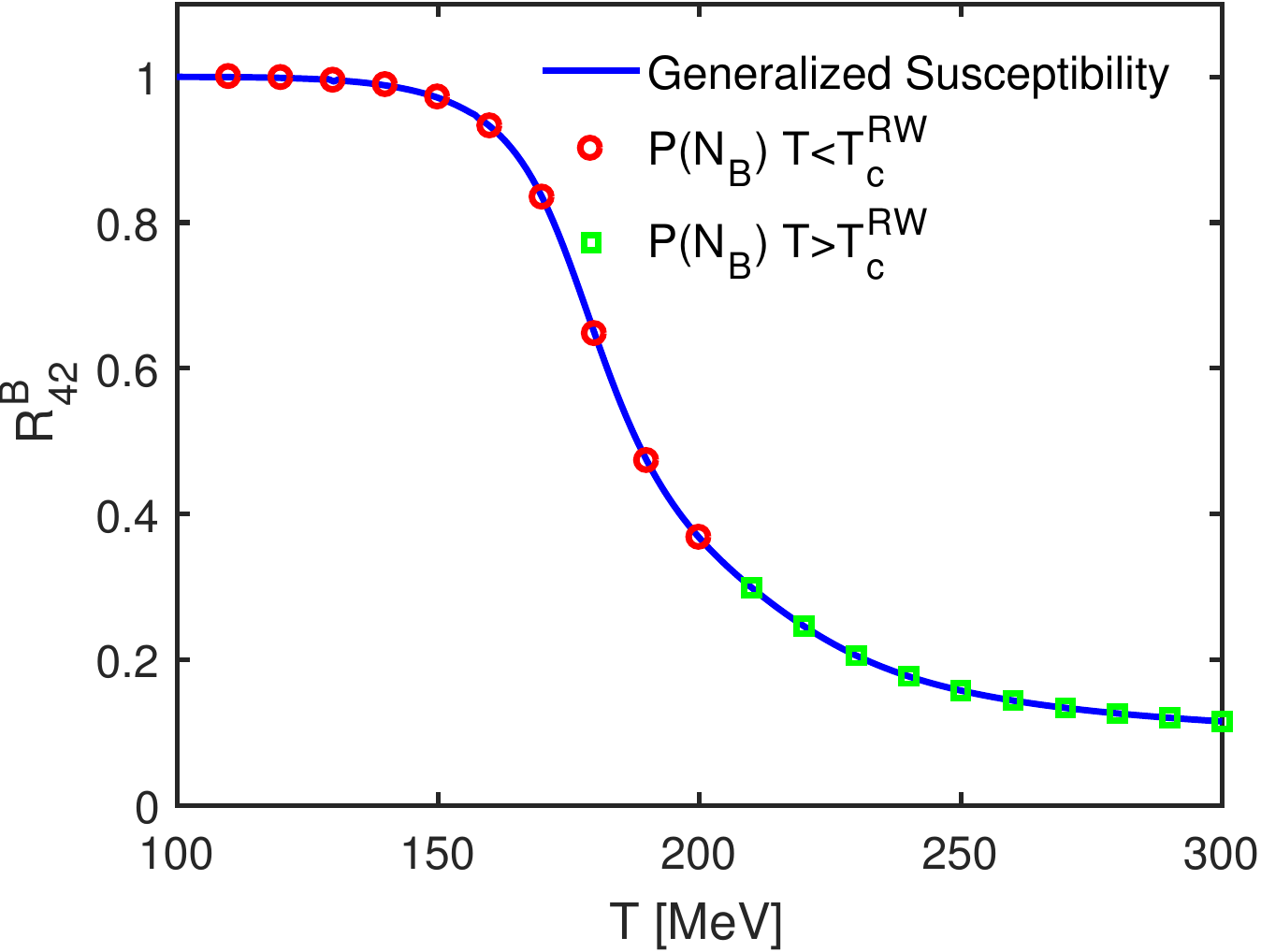}
\caption{Same as \Fig{fig:chi} but for the kurtosis of the baryon number distribution, to wit, $R^{\text{B}}_{42}=\chi^{\text{B}}_{4}/\chi^{\text{B}}_{2}$.}\label{fig:R42}
\end{figure}
%

In \Fig{fig:polyakov} we show the dependence of the modified Polyakov loop $L^{\prime}$ in \Eq{eq:Ploopmodi} on the $\theta$ with $\mu_q=i \theta T$, which is calculated in the PQM within the FRG approach at several values of the temperature. The magnitude and phase of $L^{\prime}$ are presented in different panels. One can see that the Roberge-Weiss periodicity is quite pronounced in all panels of \Fig{fig:polyakov}. We also investigate the dependence of the temperature, and confirm that when the temperature is above a critical value, that is found to be $T_c^{RW}=208\,\text{MeV}$ in our calculations, a discontinuity appears in the curve of $\sin(\phi^{\prime})$, as the solid line shows in the bottom panel of \Fig{fig:polyakov}. A recent lattice simulation by the Wuppertal-Budapest Group also find a similar value of $T_c^{RW}$ \cite{Borsanyi:2018grb}.

Figure \ref{fig:PN} shows our calculated probability distributions of the net baryon number, whose formula is given in \Eq{eq:proba}. We have discussed a lot about the canonical partition function in the numerator of \Eq{eq:proba} in former sections. The grand canonical partition function in the denominator can be easily obtained from its thermodynamic potential with a real-valued baryon chemical potential, which has been studied heavily in literatures, and we refer interested readers to such as Refs. \cite{Fu:2015naa,Fu:2015amv} etc. In \Fig{fig:PN} we only consider the case of vanishing baryon chemical potential. Therefore, the probability distributions are symmetric with respect to the zero point in $N_B$. The four subplots in \Fig{fig:PN} correspond to different values of temperature, and for every temperature, we choose two volumes. One can observe that with the increase of the volume or temperature, the profile of distribution becomes wider or ``fatter", which can be understood reasonably. We find that when the chemical potential is vanishing, the pseudo-critical temperature for the deconfinement phase transition is $T_c^{\tiny{\text{Poly}}}=178\,\text{MeV}$, and that for the chiral phase transition $T_c^{\chi}=194\,\text{MeV}$ in our calculations, which are identified by the peaks of $\partial L/\partial T$ and $\partial \rho/\partial T$, respectively.  When the temperature is low, such as $T=110\,\text{MeV}$ as shown in the top-left panel of \Fig{fig:PN}, on one hand, the chiral symmetry is broken dynamically and the constituent quark mass is large, on the other hand, quarks are confined inside baryons. Consequently, the possible net baryon number $N_B$ is restricted to a narrow region as the results of $T=110\,\text{MeV}$ show. The situation, however, is drastically changed with the increase of the temperature, especially when the temperature is above the $T_c^{\tiny{\text{Poly}}}$ and $T_c^{\chi}$. One can see that it is much easier to excite state with large $|N_B|$ in the plots of $T=200\,\text{MeV}$ as well as $T=250\,\text{MeV}$.

In \Fig{fig:compaQM} we compare the probability distribution calculated in the PQM and that in the quark-meson (QM) model, in order to illustrate 
the significance of the glue dynamics in the studies of baryon number distributions. As we discussed above, the Roberge-Weiss periodicity encoded in the glue dynamics entails that only integer $N_B$ is possible, as the blue circles show. On the contrary, loss of the periodicity, which is the case in the QM model, leads to the loss of the restriction as well. In another word, states of $N_B=N\pm 1/3$ with $N\in\mathbb{Z}$ can be excited as equally as those of $N_B=N$, as the solid dots show in \Fig{fig:compaQM}.

By employing the probability distribution of the net baryon number obtained in this work, we calculate the quadratic and quartic cumulants of the baryon number through \Eq{eq:Oaver}, and relevant results are presented in \Fig{fig:chi} with scattering symbols. In order to investigate the influence of the Roberge-Weiss phase transition, i.e., the appearance of the discontinuity in the Polyakov loop with an imaginary chemical potential when $T>T_c^{RW}$, on the physical observables $\chi^{\text{B}}_{n}$'s, we label results relevant to the temperature below and above $T_c^{RW}$ with circles and squares, respectively. Furthermore, we also show the result of $\chi^{\text{B}}_{n}$, calculated directly from the $n$-order derivative of the thermodynamic potential with respect to the real-valued baryon chemical potential, i.e., the generalized susceptibilities in \Eq{eq:suscep}, which is presented in \Fig{fig:chi} with the solid line. One can see that the two method agree with each other remarkably well. An more interesting observable is the kurtosis of the baryon number distribution, given by $R^{\text{B}}_{42}=\chi^{\text{B}}_{4}/\chi^{\text{B}}_{2}$, since it is closely related to the degree of freedom in a system. In \Fig{fig:R42} the kurtosis $R^{\text{B}}_{42}$ computed in these two methods are presented as well. As one can see, $R^{\text{B}}_{42}$ approaches to 1 in the low temperature regime, which indicates that it is the hadronic degree of freedom in the hadronic phase, which is in sharp contrast to the quark dominated system, where $R^{\text{B}}_{42}$ would have been 1/9 in the low temperature limit, for more detailed discussions, see e.g., \cite{Fu:2015naa}. Once more, we find the calculated  $R^{\text{B}}_{42}$ in these two different approaches agrees with each other very well.

Note that the agreement between these two methods is nontrivial. This is because, in the approach of the generalized susceptibilities, the thermodynamic potential with a real-valued baryon chemical potential is required. The Polyakov loop, however, is ill-defined in the SU(3) gauge theory when $\mu_B$ is real, due to the notorious sign problem. Therefore, in the actual calculations the Polyakov loop $L$ and its conjugate $\bar{L}$ are treated as two independents quantities, and their phases are ignored. Thus, one can even say that the generalized susceptibility is an approximate approach. On the contrary, when we employ the probability distribution to calculate the cumulants of the net baryon number, only an imaginary chemical potential is needed, especially when the probability distribution is symmetric with respect to $N_B=0$. Thus, the calculations in the approach of the probability distribution are exact. Another noteworthy phenomenology in \Fig{fig:chi} and \Fig{fig:R42} is that both the circles and squares coincide with the solid line very well, which indicates that the Roberge-Weiss phase transition, i.e., the discontinuity observed in the Polyakov loop when the temperature is above $T_c^{RW}$, does not affect the observables, e.g., $\chi^{\text{B}}_{2}$, $\chi^{\text{B}}_{4}$, $R^{\text{B}}_{42}$ etc.


\section{Summary and discussions}
\label{sec:sum}

In this work we investigate the probability distribution of the net baryon number in the low energy effective model within the FRG approach. Emphases are put on the influence of the glue dynamics, in particular the Roberge-Weiss periodicity, on the probability distribution of the baryon number. We find that the Roberge-Weiss periodicity directly results in that states of $N_B=N\pm 1/3$ with $N\in\mathbb{Z}$ are prohibited, and only those of $N_B=N$ are possible, which is an indication of the color confinement from another viewpoint.

By employing the probability distribution of the net baryon number obtained in our calculations, we compute the quadratic and quartic fluctuations of the baryon number, and the kurtosis of the baryon number distribution. The obtained results are compared with those from the derivatives of the thermodynamic potential with respect to the real-valued baryon chemical potential, i.e., the generalized susceptibilities. We find that these two different approaches yield consistent results.

The probability distribution of the net baryon number obtained in this work can be used as input for some transport simulations in heavy-ion collision experiments, see e.g., \cite{Xu:2016skm}, so the critical behavior near the chiral phase transition can be combined with noncritical effects, such as the  volume corrections, detector acceptance cut, resonance decays, etc., see e.g. \cite{Braun-Munzinger:2016yjz} for more discussions. All these effects are needed to be identified, before the signal of the QCD critical end point is pinned down at the BES program of RHIC \cite{Luo:2017faz}. Related work is in progress.

There is a question or problem raised in our studies. As we have discussed above, the probability distribution of the net baryon number is only possible when $N_B=N$ with $N\in\mathbb{Z}$, because of the Roberge-Weiss periodicity. It is reasonable in the hadronic phase at low temperature. The periodicity, however, is still there in the high temperature regime as shown in \Fig{fig:polyakov}; hence states of  $N_B=N\pm 1/3$ are still prohibited at high temperature, which seems that it is not consistent with the picture of deconfined quarks. Another more amazing finding is that, even with the hadronic probability distribution of the net baryon number in the high temperature limit, the predicted cumulants of the baryon number distributions in \Fig{fig:chi} and \Fig{fig:R42} are consistent with those obtained from the approach of generalized susceptibilities, which is exotic. More studies are needed to answer these questions in the future.

\begin{acknowledgments}

The work was supported by the National Natural Science Foundation of China under Contracts Nos. 11775041.

\end{acknowledgments}


\bibliography{ref-lib.bib}

\end{document}